\documentclass[aps,prd,twocolumn,floatfix]{revtex4-1}

\usepackage{graphicx}
\usepackage{epsfig}

\newcommand{\be}{\begin{equation}}
\newcommand{\ee}{\end{equation}}
\newcommand{\ba}{\begin{eqnarray}}
\newcommand{\ea}{\end{eqnarray}}

\begin{document}

\title{A Metric for Rapidly Spinning Black Holes Suitable for Strong-Field Tests of the No-Hair Theorem}

\author{Tim Johannsen and Dimitrios Psaltis}
\affiliation{Physics and Astronomy Departments, University of Arizona, 1118 E. 4th Street, Tucson, AZ 85721, USA}

\date{\today}

\begin{abstract}

According to the no-hair theorem, astrophysical black holes are uniquely characterized by their masses and spins and are described by the Kerr metric. Several parametric deviations from the Kerr metric have been suggested to study observational signatures in both the electromagnetic and gravitational-wave spectra that differ from the expected Kerr signals. Due to the no-hair theorem, however, such spacetimes cannot be regular everywhere outside the event horizons, if they are solutions to the Einstein field equations; they are often characterized by naked singularities or closed time-like loops in the regions of the spacetime that are accessible to an external observer. For observational tests of the no-hair theorem that involve phenomena in the vicinity of the circular photon orbit or the innermost stable circular orbit around a black hole, these pathologies limit the applicability of the metrics only to compact objects that do not spin rapidly. In this paper, we construct a Kerr-like metric which depends on a set of free parameters in addition to its mass and spin and which is regular everywhere outside of the event horizon. We derive expressions for the energy and angular momentum of a particle on a circular equatorial orbit around the black hole and compute the locations of the innermost stable circular orbit and the circular photon orbit. We demonstrate that these orbits change significantly for even moderate deviations from the Kerr metric. The properties of our metric make it an ideally suited spacetime to carry out strong-field tests of the no-hair theorem in the electromagnetic spectrum using the properties of accretion flows around astrophysical black holes of arbitrary spin.

\end{abstract}

\maketitle

\section{Introduction}

The no-hair theorem encapsulates the remarkable property of general-relativistic black holes that these objects are fully and uniquely characterized by their masses and spins and are described by the Kerr metric. According to the no-hair theorem, the Kerr metric is the only stationary, axisymmetric, asymptotically flat vacuum spacetime in general relativity that has an event horizon but no closed timelike curves outside of the horizon \cite{NHT,Carter73}. Mass $M$ and spin $J$ are the first two (Geroch-Hansen) multipole moments of the Kerr spacetime, and all higher order moments can be expressed in terms of these two moments. The multipole moments consist of a set of mass multipole moments $M_l$, which vanish if $l$ is odd, and a set of current multipole moments $S_l$, which vanish if $l$ is even. The no-hair theorem can then be expressed by the relation \cite{Multipoles}
\be
M_l + {\rm i}S_l = M({\rm i}a)^l,
\ee
where $a\equiv J/M$ is the spin parameter.

Despite a wealth of observational evidence for the existence of astrophysical black holes (see discussion in, e.g., \cite{Psaltis06}), a definite proof is still lacking. Several potential tests of the no-hair theorem have been suggested using observations of gravitational waves from extreme mass-ratio inspirals \cite{EMRIs,BC07,CH04,GB06,Gair08,VH10} and observations in the electromagnetic spectrum of accreting black holes \cite{PaperI,PaperII,PaperIII,EM}, of stars on an orbit around Sgr A* \cite{Will08,Merritt10}, and of pulsar black-hole binaries \cite{WK99}. For recent reviews, see \cite{Reviews}.

Observational tests of this kind require a framework that is based on spacetimes that deviate from the Kerr metric by one or more parameters (e.g., \cite{MN92,CH04,GB06,VH10,Vigeland11}). These spacetimes have a modified multipole structure that is given by a relation of the form \cite{CH04,VH10}
\be
M_l + {\rm i}S_l = M({\rm i}a)^l + \delta M_l + {\rm i}\delta S_l
\label{deltamult}
\ee
with deviations $\delta M_l$ and $\delta S_l$.

Parametric deviations of the form (\ref{deltamult}) harbor a compact object that is a general-relativistic black hole only if all corrections $\delta M_l$ and $\delta S_l$ are equal to zero. Within general relativity, measuring these parametric deviations constitutes a null-test that investigates the nature of compact objects \cite{CH04,Hughes06}. General relativity, however, has been marginally tested in the regime of strong gravitational fields (e.g., \cite{PsaltisLRR}), and astrophysical black holes might not be Kerr black holes as predicted by the no-hair theorem \cite{PaperI} (see, also, \cite{YP09,SY09}).

Because of the no-hair theorem, all parametric deviations of the Kerr metric within general relativity have to violate at least one of the prerequisites of this theorem. Consequently, these spacetimes contain either singularities or regions with closed timelike curves outside of the event horizon.  The degree to which these pathologies affect different proposed tests of the no-hair theorem depends on the intended application. For all the currently proposed metrics that deviate from the Kerr solution, pathologies appear very close to the corresponding Kerr event horizon \cite{JP11b}. As a result, they do not hamper tests of the no-hair theorem that involve the orbits of objects at large distances from the horizons, as is the case, e.g., for test with extreme mass-ratio inspirals \cite{BC07} or observations of stars and pulsars around black holes \cite{Will08,WK99}. These pathologies, however, become prohibitive in cases of tests that involve observations of the images of the inner accretion flows \cite{PaperI,PaperII} or of X-ray observations of quasi-periodic oscillations, of fluorescent iron lines, or of the continuum spectra of accretion disks \cite{PaperI,PaperIII}.

For the latter tests and for moderately spinning black holes, the singularities or closed time-like loops appear far inside the location of the photon orbit and the location of the ISCO, both of which dominate the observational characteristics of black holes. These pathologies can, therefore, be handled by imposing an artificial cutoff with an inflow boundary condition at some radius in the exterior spacetime, between the location of the pathologies and the location of the photon orbit or ISCO. For rapidly spinning black holes, however, the radius of the ISCO becomes comparable to the radius of the horizon and imposing such an artificial cut-off is no longer possible. This limits the applicability of current parametric deviations of the Kerr metric for several observational tests of the no-hair theorem in the electromagnetic spectrum to only moderately spinning black holes \cite{JP11b}.

Performing tests of the no-hair theorem with observations of phenomena that occur in the vicinity of the circular photon orbit or the ISCO around a black holes requires that we use a metric that is free of such pathologies for arbitrary values of the spin. However, finding a metric of this kind is a highly nontrivial task. Introducing small parametric deviations to individual elements of the metric in an arbitrary manner routinely leads to the pathologies discussed above. In order for such a spacetime to describe a black hole, it can no longer be a solution of the Einstein equations, because otherwise it would render the no-hair theorem false. To date, however, black hole metrics for theories that obey the Einstein equivalence principle \cite{Will93} are only known for static black holes (e.g., \cite{YS11,DS10}), for slowly rotating black holes with parity violations \cite{YP09}, or in Einstein-Dilaton-Gauss-Bonnet gravity \cite{PC09}.

In this paper, we construct such a Kerr-like black hole metric which suffers from no pathologies up to the maximum value of the spin and which contains a set of parameters that measure potential deviations from the Kerr metric in the strong-field regime. In order to achieve this in a regular manner, we start by introducing a parametric deviation to the Schwarzschild metric, following Ref. \cite{YS11}. We then apply the Newman-Janis algorithm \cite{NJalgorithm}, as in Ref. \cite{VH10}, in order to generate a metric for a spinning black hole.

We take special care to retain several properties that make the Kerr metric unique in performing ray-tracing calculations in general relativity. Our metric shares the same non-zero metric elements with the Kerr solution, which allows for a straightforward implementation for calculations of ray tracing with existing geodesic algorithms and an intuitive interpretation of observables.  We likewise obtain constraints for some of the parameters of our metric from observational limits on modifications of general relativity in the weak-field regime as well as from the requirements of asymptotic flatness.

For the particular case of only one deviation parameter, we show that our metric is regular everywhere outside the horizon for the entire range of allowable spins up to a maximum value, which depends on the deviation. It can, therefore, be used to study astrophysical phenomena arbitrarily close to the event horizon and to test the no-hair theorem in the electromagnetic spectrum even with rapidly spinning black holes. We also derive expressions for the energy and angular momentum of a particle on a circular equatorial orbit around the central black hole and compute the locations of the innermost stable circular orbit (ISCO) and the circular photon orbit as a function of spin and the deviation parameter.

In Section~2, we construct our new metric. We constrain the set of free parameters in Section~3 and analyze the properties of our metric in Section~4. We summarize our conclusions in Section~5.

\section{Construction of a Kerr-like Black Hole Metric}

In this section, we construct a new class of Kerr-like black hole metrics which describe a stationary, axisymmetric, and asymptotically flat vacuum spacetime. In addition to the mass and spin of the black hole, this spacetime depends on a set of parameters that measure potential deviations from the Kerr metric. Our spacetime reduces smoothly to the Kerr metric if the deviations are dialed to zero.

Our starting point is a Schwarzschild-like metric with the line element \cite{YS11}
\begin{eqnarray}
ds^2 &=& -f [1+h(r)] dt^2 + f^{-1} [1+h(r)] dr^2 \nonumber \\
&+& r^2 (d\theta^2 + \sin^2\theta d\phi^2)
\label{SchwSchw}
\end{eqnarray}
in Schwarzschild coordinates $(t,r,\theta,\phi)$, where $M$ is the mass of the central object and
\be
f \equiv 1-\frac{2M}{r}.
\ee

A metric of this form is both stationary and spherically symmetric and reduces to the Schwarzschild metric in the case $h(r)=0$. As in \cite{YS11}, we do not modify the angular part of the metric for simplicity and in order to retain spherical symmetry. Unlike \cite{GB06,VH10}, since we interested in constructing a black hole spacetime that is free of pathologies outside of the event horizon, we do not require our metric to be a vacuum solution of the Einstein equations. Similarly, we do not require full integrability of geodesic motion in our metric (unlike \cite{Vigeland11}). While this property is critical for the design of waveforms for observations in the gravitational-wave spectrum, it may only simplify ray-tracing calculations for applications in the electromagnetic spectrum, but it is not a necessity.

We choose the function $h(r)$ to be of the form
\be
h(r) \equiv \sum_{k=0}^\infty \epsilon_k \left( \frac{M}{r} \right)^k.
\label{h(r)}
\ee

The Kerr metric can be obtained from the Schwarzschild metric by the Newman-Janis algorithm \cite{NJalgorithm}, which is based on a complex coordinate transformation. Through this procedure the effect of rotation can be incorporated into a static spacetime in a natural way. In the following, we apply the Newman-Janis algorithm to the Schwarzschild-like metric given by Eq. (\ref{SchwSchw}) in order to construct a Kerr-like metric that depends on the mass $M$, spin $a$, and the set of parameters $\epsilon_k$.

First we perform a transformation to Eddington-Finkelstein coordinates choosing a set of new coordinates $(u',r',\theta',\phi')$, where
\ba
u' &=& t-r-2M\ln\left( \frac{r-2M}{2M} \right), \\
r' &=& r,~~~\theta' = \theta,~~~\phi' = \phi,
\ea
which yields a metric $\tilde{g}_{\mu\nu}$ of the form
\begin{eqnarray}
ds^2 &=& -f [1+h(r)] du^2 - 2[1+h(r)]dudr \nonumber \\
&+& r^2 (d\theta^2 + \sin^2\theta d\phi^2).
\label{SchwEF}
\end{eqnarray}
In expression (\ref{SchwEF}) we have dropped the primes for brevity.

We express the metric $\tilde{g}_{\mu\nu}$ given by Eq. (\ref{SchwEF}) in contravariant form in the Newman-Penrose formalism \cite{NPformalism}
\begin{equation}
\tilde{g}^{\mu\nu} = -l^\mu n^\nu - l^\nu n^\mu + m^\mu \bar{m}^\nu + m^\nu \bar{m}^\mu
\label{NPmetric}
\end{equation}
using a complex null tetrad
\begin{equation}
Z^\mu_s = (l^\mu,n^\mu,m^\mu,\bar{m}^\mu),~~~~~s = 1,2,3,4
\end{equation}
with legs
\begin{eqnarray}
l^\mu &=& \delta^\mu_r, \\
n^\mu &=& \frac{1}{1+h(r)} \left[ \delta^\mu_u - \frac{1}{2} \left(1-\frac{2M}{r}\right) \delta^\mu_r \right], \\
m^\mu &=& \frac{1}{\sqrt{2}r} \left( \delta^\mu_\theta + \frac{i}{\sin\theta} \delta^\mu_\phi \right), \\
\bar{m}^\mu &=& \frac{1}{\sqrt{2}r} \left( \delta^\mu_\theta - \frac{i}{\sin\theta} \delta^\mu_\phi \right).
\end{eqnarray}
This tetrad is orthonormal obeying the conditions
\ba
&&l_\mu m^\mu = \l_\mu \bar{m}^\mu = n_\mu m^\mu = n_\mu \bar{m}^\mu = 0, \\
&&l_\mu l^\mu = n_\mu n^\mu = m_\mu m^\mu = \bar{m}_\mu \bar{m}^\mu = 0, \\
&&l_\mu n^\mu = -1,~~~ m_\mu \bar{m}^\mu = 1.
\ea

Now we allow for the radius $r$ to take on complex values and rewrite the legs of the null tetrad in the form
\ba
l^\mu &=& \delta^\mu_r, \\
n^\mu &=& \frac{1}{1+h(r,\bar{r})} \left[ \delta^\mu_u - \frac{1}{2} \left(1-\frac{M}{r}-\frac{M}{\bar{r}}\right) \delta^\mu_r \right], \\
m^\mu &=& \frac{1}{\sqrt{2}r} \left( \delta^\mu_\theta + \frac{i}{\sin\theta} \delta^\mu_\phi \right), \\
\bar{m}^\mu &=& \frac{1}{\sqrt{2}\bar{r}} \left( \delta^\mu_\theta - \frac{i}{\sin\theta} \delta^\mu_\phi \right),
\ea
where an overbar denotes complex conjugation and
\be
h(r,\bar{r}) \equiv \sum_{k=0}^\infty \left[ \epsilon_{2k} + \epsilon_{2k+1} \frac{M}{2} \left( \frac{1}{r} + \frac{1}{\bar{r}} \right) \right] \left( \frac{M^2}{r\bar{r}} \right)^{k}.
\ee

Next we perform a complex coordinate transformation defining a new set of coordinates $(u',r',\theta',\phi')$ by the relations
\ba
u' &=& u - ia\cos\theta, \\
r' &=& r + ia\cos\theta, \\
\theta' &=& \theta,~~~\phi' = \phi.
\ea
We transform the tetrad in the usual way,
\be
Z'^\mu_s = \frac{\partial x'^\mu}{\partial x^\nu} Z^\nu_s,
\ee
and obtain
\ba
l^\mu &=& \delta^\mu_r, \\
n^\mu &=& \frac{1}{1+h(r,\bar{r})} \left[ \delta^\mu_u - \frac{1}{2} \left(1-\frac{2Mr}{\Sigma}\right) \delta^\mu_r \right], \\
m^\mu &=& \frac{1}{\sqrt{2}r} \left[ ia\sin\theta \left(\delta^\mu_u - \delta^\mu_r \right) + \delta^\mu_\theta + \frac{i}{\sin\theta} \delta^\mu_\phi \right], \\
\bar{m}^\mu &=& \frac{1}{\sqrt{2}\bar{r}} \left[ -ia\sin\theta \left(\delta^\mu_u - \delta^\mu_r \right) + \delta^\mu_\theta - \frac{i}{\sin\theta} \delta^\mu_\phi \right],
\ea
where
\ba
\Sigma &\equiv& r^2 + a^2 \cos^2\theta, \\
h(r,\theta) &\equiv& \sum_{k=0}^\infty \left( \epsilon_{2k} + \epsilon_{2k+1}\frac{Mr}{\Sigma} \right) \left( \frac{M^2}{\Sigma} \right)^{k},
\label{h(r,theta)}
\ea
and, again, we have dropped the primes.

From these expressions, we recover the contravariant metric with the use of Eq. (\ref{NPmetric}) and perform a transformation to coordinates $(t',r',\theta',\phi')$ given by the implicit relations
\ba
du &=& dt' + \frac{r'^2 + a^2}{\Delta'}dr', \\
r &=& r',~~~\theta = \theta', \\
d\phi &=& d\phi' - \frac{a}{\Delta'}dr',
\ea
where
\be
\Delta \equiv r^2 - 2Mr + a^2.
\ee

In the case that the function $h(r,\theta)$ vanishes, the metric derived in this fashion is the usual Kerr metric in Boyer-Lindquist coordinates with mass $M$ and spin $a$. For nonzero values of the function $h(r,\theta)$, however, the resulting metric contains the off-diagonal element
\be
\tilde{g}_{r\phi} = \frac{ a \Sigma \sin^2\theta }{\Delta} h(r,\theta)
\ee
in addition to the usual frame-dragging element $\tilde{g}_{t\phi}$.

In order to eliminate the element $\tilde{g}_{r\phi}$, we apply another transformation to new coordinates $(t',r',\theta',\phi')$ given by the implicit relations
\ba
dt &=& dt' + F(r',\theta')dr', \\
r &=& r', \\
\theta &=& \theta', \\
d\phi &=& d\phi' + G(r',\theta')dr'
\ea
with the functions
\ba
F(r',\theta') \equiv -\frac{ \tilde{g}_{r\phi} }{ \tilde{g}_{tt} } \left( \frac{ \tilde{g}_{t\phi} }{ \tilde{g}_{tt} } - \frac{ \tilde{g}_{\phi\phi} }{ \tilde{g}_{t\phi} } \right)^{-1}, \\
G(r',\theta') \equiv \frac{ \tilde{g}_{r\phi} }{ \tilde{g}_{t\phi} } \left( \frac{ \tilde{g}_{t\phi} }{ \tilde{g}_{tt} } - \frac{ \tilde{g}_{\phi\phi} }{ \tilde{g}_{t\phi} } \right)^{-1}.
\ea

Finally (dropping the primes), we arrive at the following metric $g_{\mu\nu}$ given by the line element
\begin{widetext}
\ba
ds^2 = && -[1+h(r,\theta)] \left(1-\frac{2Mr}{\Sigma}\right)dt^2 -\frac{ 4aMr\sin^2\theta }{ \Sigma }[1+h(r,\theta)]dtd\phi + \frac{ \Sigma[1+h(r,\theta)] }{ \Delta + a^2\sin^2\theta h(r,\theta) }dr^2 + \Sigma d\theta^2 \nonumber \\
&& + \left[ \sin^2\theta \left( r^2 + a^2 + \frac{ 2a^2 Mr\sin^2\theta }{\Sigma} \right) + h(r,\theta) \frac{a^2(\Sigma + 2Mr)\sin^4\theta }{\Sigma} \right] d\phi^2,
\label{metric}
\ea
\end{widetext}
which reduces to the Kerr metric in Boyer-Lindquist coordinates in the case $h(r,\theta)=0$ and to the generalized Schwarzschild metric given by Eq. (\ref{SchwSchw}) if $a=0$.

The metric $g_{\mu\nu}$ that we have constructed in this manner is both stationary and axisymmetric. As we will argue in the following, the nontrivial dependence of our metric on the function $h(r,\theta)$ ensures the preservation of the properties of the Kerr metric that are critical for observational tests of the no-hair theorem. In general relativity, the Einstein tensor of our metric is nonzero unless $h(r,\theta)$ vanishes. Therefore, we regard our metric as a vacuum spacetime of an appropriately chosen set of field equations which are unknown but different from the Einstein equations for nonzero $h(r,\theta)$. For observational tests of the no-hair theorem in the electromagnetic spectrum, the field equations are not needed explicitly \cite{Psaltis09}, and we only require a spacetime and the validity of the Einstein equivalence principle (c.f., \cite{Will93}), which governs the motion of particles in that spacetime.

We justify the nature of our metric in Section~4, where we show that its properties are very similar to the ones of the Kerr metric. In particular, we compute the location of the event horizon. The requirement of asymptotic flatness imposes restrictions on the function $h(r,\theta)$, which we will address in the next section.

\section{Constraints on the Function $h(r,\theta)$}

\begin{figure*}[ht]
\begin{center}
\psfig{figure=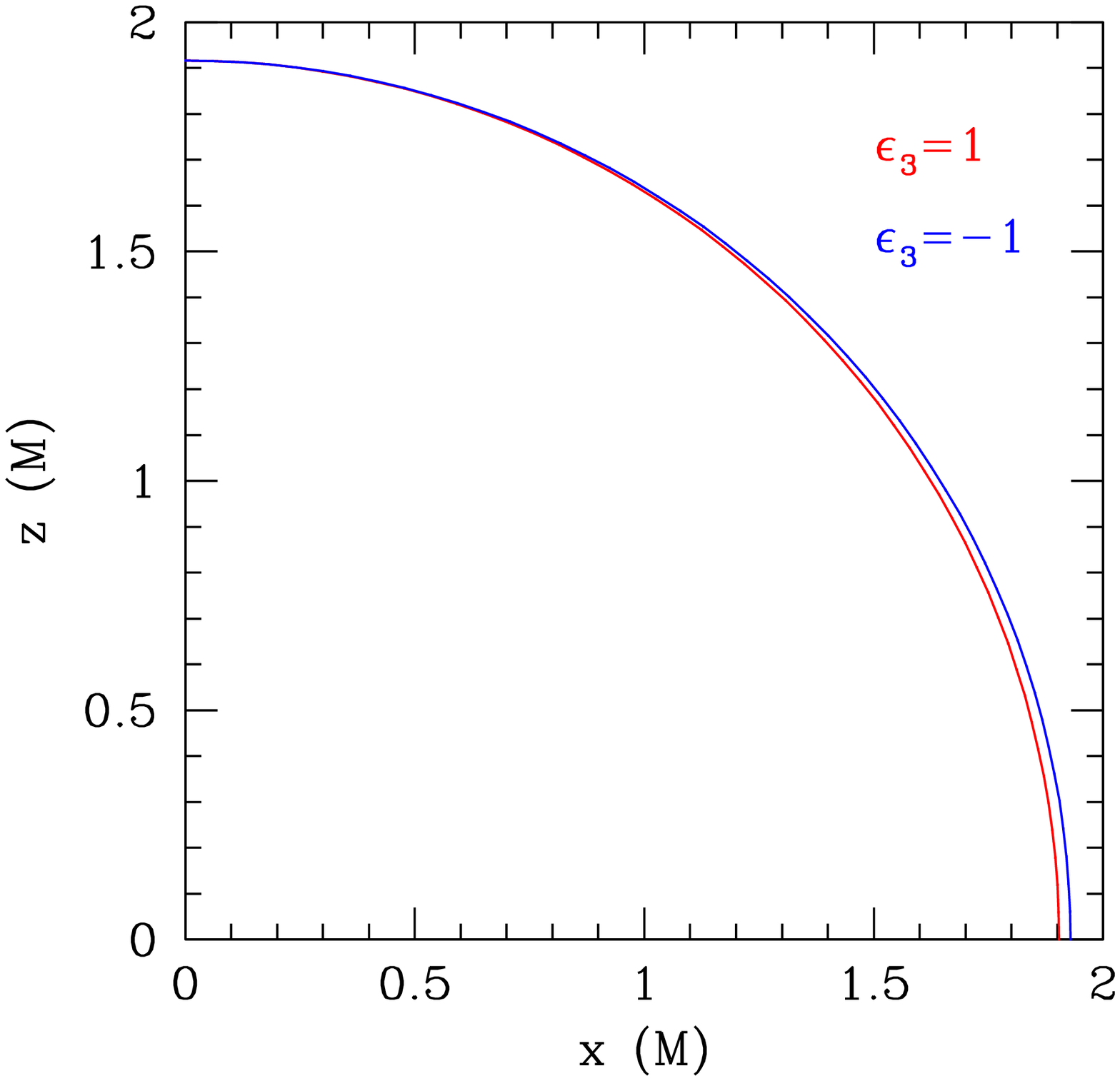,height=2.3in}
\psfig{figure=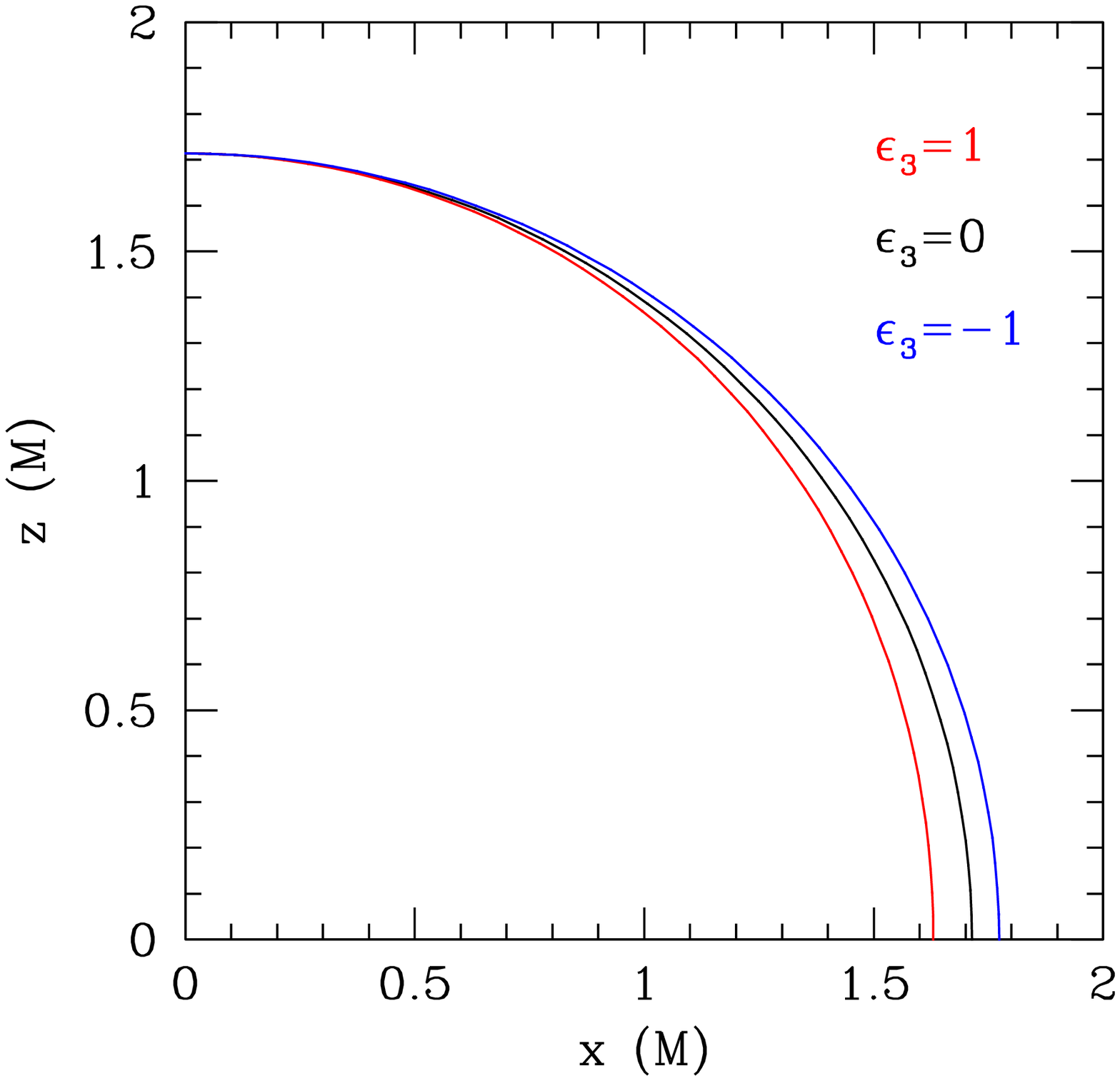,height=2.3in}
\psfig{figure=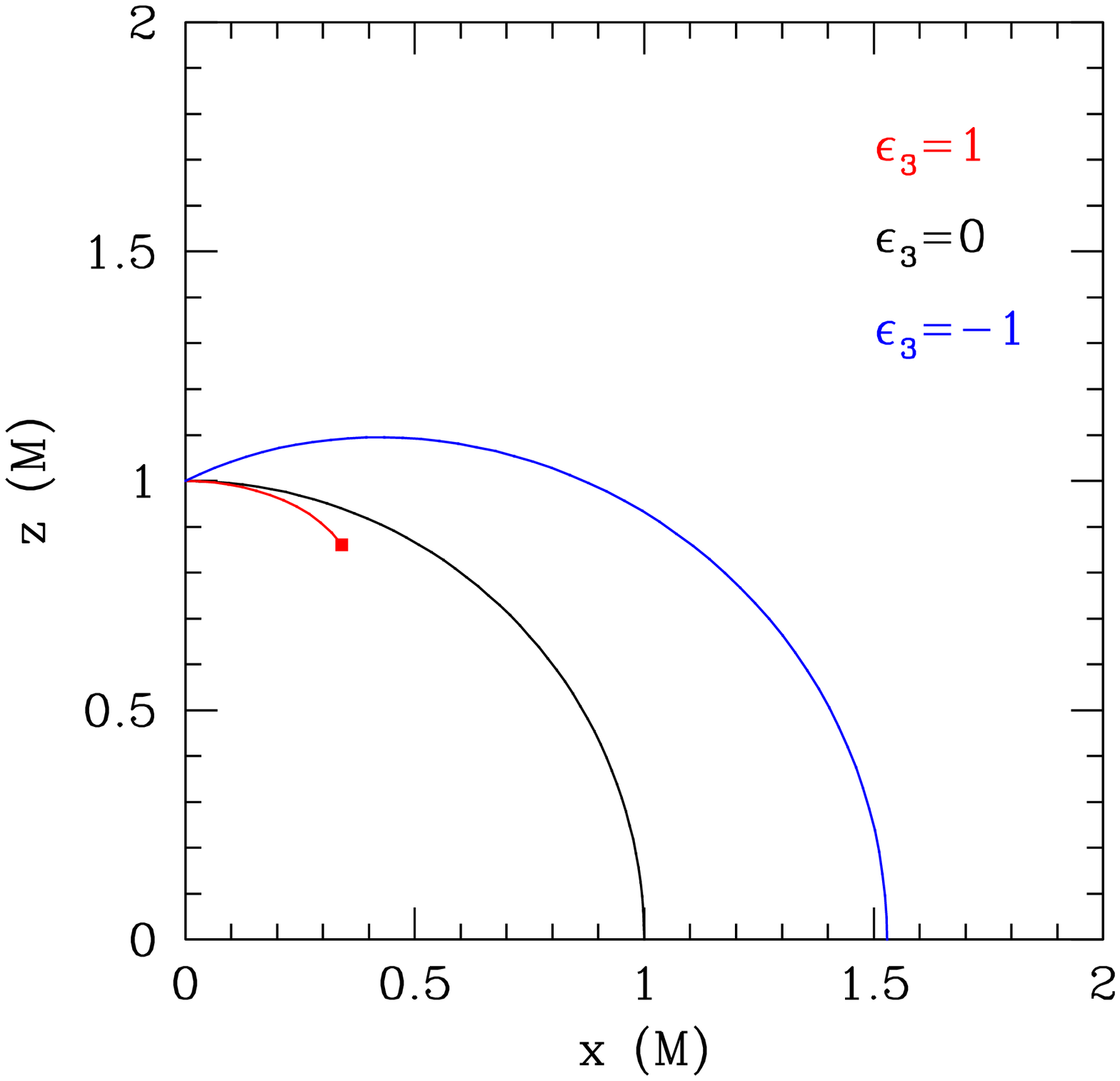,height=2.3in}
\end{center}
\caption{Event horizon of a non-Kerr black hole in the $xz$-plane for values of the spin (left) $a=0.4M$, (middle) $a=0.7M$, and (right) $a=1.0M$ and several values of the parameter $\epsilon_3$. For positive values of the parameter $\epsilon_3$, the event horizon has a more prolate shape than the horizon of a Kerr black hole with the same spin, while for negative values of the parameter $\epsilon_3$, the shape is more oblate. In the case $a=M$, the event horizon has a dumbbell shape if $\epsilon_3=-1$ and is not closed if $\epsilon_3=1$.} 
\label{horizon}
\end{figure*}

In this section, we constrain the form of the function $h(r,\theta)$ given by Eq. (\ref{h(r,theta)}) by the requirements that the metric $g_{\mu\nu}$ given by Eq. (\ref{metric}) is asymptotically flat and consistent with observational weak-field constraints on deviations from the Kerr metric. The resulting metric, then, is suitable for the exploration of the strong-field regime in the vicinity of black holes.

In Newtonian gravity and at a large distance from the source, the potential of an extended body approaches that of a spherical body of equal mass. Similarly, in general relativity, stationary and asymptotically flat spacetimes are Schwarzschild-like in the limit of large radii in an appropriately chosen coordinate system, i.e., they fall off as $1/r$ or faster \cite{Beig}. In that particular gauge, such metrics are of the asymptotic form (e.g., \cite{Heusler1996})
\ba
ds^2 = &-&\left[ 1-\frac{2M}{r} + \mathcal{O}\left(r^{-2}\right) \right]dt^2 \nonumber \\
&-& \left[\frac{4a}{r}\sin^2\theta + \mathcal{O}\left(r^{-2}\right) \right]dtd\phi \nonumber \\
&+& \bigg[1 + \mathcal{O}\left(r^{-1}\right) \bigg] \bigg[dr^2 + r^2 d\Omega^2 \bigg],
\label{asympt}
\ea
where we used the notation
\be
d\Omega^2 = d\theta^2 + \sin^2\theta d\phi^2.
\ee
Asymptotically flat spacetimes with a slower fall-off involve gravitational radiation \cite{Kennefick1995} and, thus, cannot be stationary.

A similar argument must hold for more general spacetimes that are not necessarily a solution of the Einstein equations. For $r\gg M$ and $r\gg a$, our metric given by Eq. (\ref{metric}) has the asymptotic form
\ba
ds^2 \approx &-&\left[ 1 - \frac{2M}{r} + h(r) \right] dt^2 \nonumber \\
&-& \frac{4a[1+h(r)]}{r}\sin^2\theta dtd\phi \nonumber \\
&+& \left[ 1 + \frac{2M}{r} + h(r) \right] dr^2 + r^2 d\Omega,
\label{asymptmetric}
\ea
where $h(r)$ is given by Eq. (\ref{h(r)}). Therefore, the function $h(r)$ must be of order $\mathcal{O}(1/r^n)$ with $n\geq2$, and we conclude that $\epsilon_0=\epsilon_1=0$.

Limits on the parameter $\epsilon_2$ of the next leading-order term in the function $h(r)$ can readily be obtained from the observational constraints on weak-field deviations from general relativity in the parameterized post-Newtonian (PPN) framework \cite{WillLivRev}. In the PPN approach, the asymptotic spacetime is expressed as
\be
ds^2 = -A(r)dt^2 + B(r)dr^2 + r^2 d\Omega,
\ee
where
\ba
A(r) &=& 1 - \frac{2M}{r} + 2(\beta - \gamma)\frac{M^2}{r^2}, \\
B(r) &=& 1 + 2\gamma \frac{M}{r}.
\ea
In general relativity, $\beta=\gamma=1$.

From the asymptotic form of the metric given by Eq. (\ref{asymptmetric}), we identify
\ba
&& \epsilon_2 = 2(\beta-1), \\
&& \gamma = 1.
\ea
The best current PPN contraint on the parameter $\beta$ is set by the Lunar Laser Ranging experiment and yields \cite{LLR}
\be
|\beta - 1| \leq 2.3 \times 10^{-4},
\ee
if the weak equivalence principle is satisfied, which we assume throughout the paper. Therefore, this limit implies that
\be
|\epsilon_2| \leq 4.6 \times 10^{-4}.
\ee

For the remainder of this paper, we will set $\epsilon_2=0$ and explore in some detail metrics with $\epsilon_k=0$ for $k>3$. In this case, the function $h(r,\theta)$ reduces to
\be
h(r,\theta) = \epsilon_3 \frac{M^3 r}{\Sigma^2}.
\label{hchoice}
\ee
The parameter $\epsilon_3$ is unconstrained by current observational tests of general relativity (c.f. \cite{PsaltisLRR}). Our metric with this choice of $h(r,\theta)$, therefore, allows us to probe the regime of strong-field gravity in parametric form.

\section{Metric Properties}

In this section, we analyze some of the properties of the metric given by Eq. (\ref{metric}), and we choose the function $h(r,\theta)$ according to Eq. (\ref{hchoice}) for simplicity. In particular, we determine the range of the parameters $a$ and $\epsilon_3$ for which our metric describes a black hole. A similar analysis should be valid for all higher orders in $M/r$. Unless the parameter $|\epsilon_3|$ is very small, we expect potential strong-field deviations from the Kerr metric to be most easily detectable at order $(M/r)^3$.

\subsection{Event Horizon}

First, we calculate the location of the event horizon, which occurs at the root of the equation
\be
g_{t\phi}^2 - g_{tt} g_{\phi\phi} = 0.
\label{eq:horizon}
\ee
This equation can be rewritten in the form
\be
\left( 1 + \epsilon_3 \frac{M^3 r}{\Sigma^2} \right) w(r,\theta;M,a,\epsilon_3) = 0,
\label{eq:horizon2}
\ee
where $w(r,\theta;M,a,\epsilon_3)$ is a function of the radius $r$ and the angle $\theta$, as well as of the mass $M$, spin $a$, and the parameter $\epsilon_3$. This equation can have more than one root leading to the presence of both an inner and an outer horizon similar to the case of the Kerr metric. Since in this paper we are only concerned with the exterior spacetime, we will refer hereafter to the outer horizon simply as the event horizon.

In Figure~\ref{horizon}, we plot the event horizon in the $xz$-plane for several values of the spin $a$ and the parameter $\epsilon_3$. The horizon is more prolate than the horizon of a Kerr black hole with the same spin for positive values of the parameter $\epsilon_3$, while it is more oblate for negative values of the parameter $\epsilon_3$. In the case $a=M$, the event horizon has a dumbbell shape if $\epsilon_3=-1$ and is not closed if $\epsilon_3=1$.

In this paper, we are only interested in black holes, i.e., in compact objects for which the event horizon is entirely closed. In the case $a=0$, the event horizon is a sphere with radius $r_{\rm h}=2M$, if $\epsilon_3\geq-8$, or $r_{\rm h}=(|\epsilon_3|)^{1/3}M$, if $\epsilon_3<-8$. For negative values of the parameter $\epsilon_3$, the event horizon is always closed, because the first factor in Eq. (\ref{eq:horizon2}) vanishes at some $r>0$ for all $0\leq\theta<\pi$. For positive values of the parameter $\epsilon_3$, the first factor in Eq. (\ref{eq:horizon2}) is always positive, while the existence of a root of the function $w(r,\theta;M,a,\epsilon_3)$ depends on the value of the parameter $\epsilon_3$. For each value of the spin $|a|>0$, there exists a value of the parameter $\epsilon_3>0$ such that the event horizon is no longer closed. A hole appears in the event horizon around the equatorial plane within the range $\theta=\pi/2\pm\theta_{\rm hole}$, and the central object becomes a naked singularity.

\begin{figure}[ht]
\begin{center}
\psfig{figure=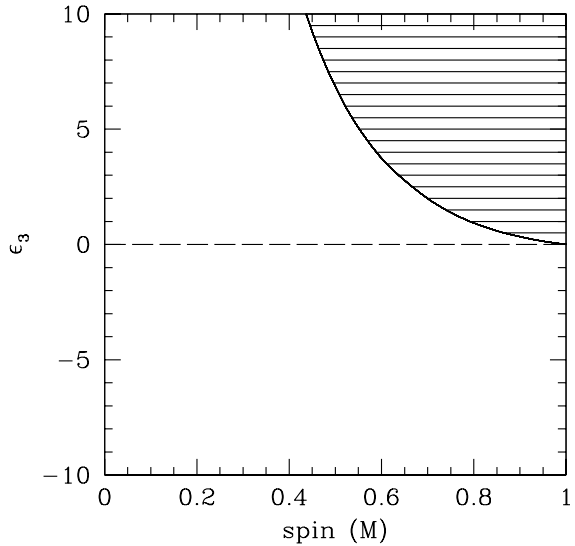,height=3.3in}
\end{center}
\caption{Maximum values of the parameter $\epsilon_3$ versus the spin $a$, for which the event horizon is entirely closed. The shaded region marks the part of the parameter space where the central object is a naked singularity. Outside of this region, the central object is a black hole, which is described by its mass $M$, spin $a$ and the parameter $\epsilon_3$. The dashed line corresponds to a Kerr black hole.}
\label{f:maxdat}
\end{figure}

In Figure~\ref{f:maxdat}, we delineate the part of the parameter space, within which the event horizon is closed, and the central object is a black hole. The solid line marks the upper limit on the parameter $\epsilon_3$ as a function of the spin, for which the event horizon is still closed. The shaded region corresponds to the excluded part of the parameter space, where the central object is a naked singularity. In principle, the parameter space can be expanded to include values of the parameter $|\epsilon_3|>10$. However, we will not consider this case here, because this relatively large range of the parameter $\epsilon_3$ should already suffice to study strong-field deviations from the Kerr metric.

The Kerr metric describes a black hole only for values of the spin $|a|\leq M$. In the case $|a|>M$, this spacetime contains a naked singularity, and causality is violated at every point in space due to the presence of closed timelike curves \cite{Carter68,Carter73}. In our metric, the event horizon is not closed if $|a|>M$, unless $\epsilon_3 < -16|a|^3 / 3\sqrt{3}$. In this case, our metric describes a superspinning black hole (c.f. \cite{BF09}). We will not consider this case either, because the Kerr metric itself is unphysical in this spin range.

Analyzing the elements of our metric, we find that $g_{\theta\theta}>0$ throughout the spacetime and $g_{rr}>0$, $g_{\phi\phi}>0$ outside of the event horizon. Consequently, our (exterior) spacetime is free of closed timelike curves, and causality is satisfied.

\begin{figure}[ht]
\begin{center}
\psfig{figure=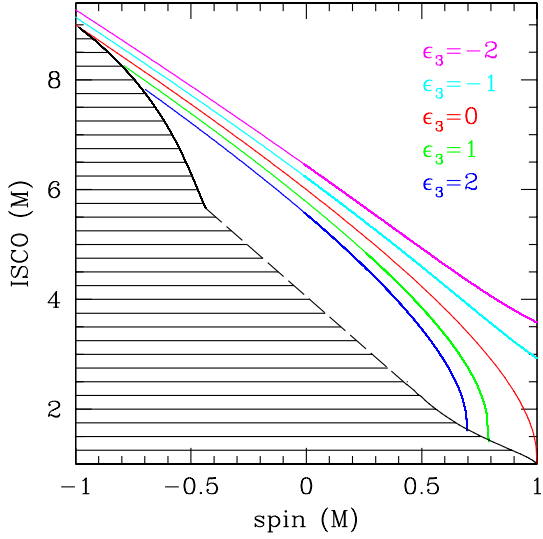,height=3.3in}
\end{center}
\caption{Radius of the ISCO as a function of the spin $a$ for several values of the parameter $\epsilon_3$. The radius of the ISCO decreases with increasing values of the parameter $\epsilon_3$. The shaded region marks the excluded part of the parameter space.} 
\label{f:isco}
\end{figure}

\begin{figure}[ht]
\begin{center}
\psfig{figure=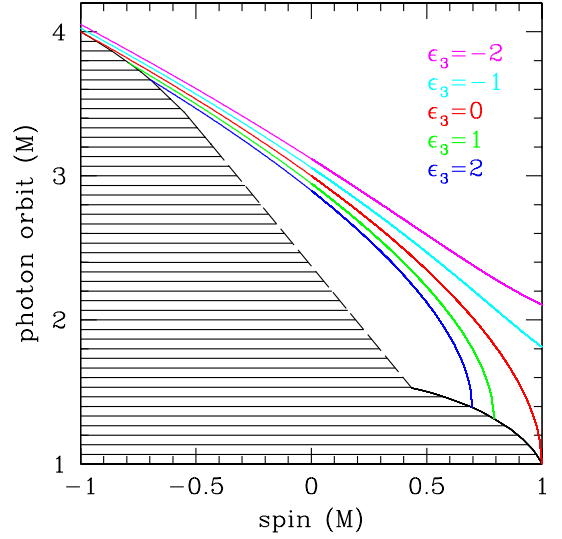,height=3.3in}
\end{center}
\caption{Radius of the circular photon orbit as a function of the spin $a$ for several values of the parameter $\epsilon_3$. The radius of the circular photon decreases with increasing values of the parameter $\epsilon_3$. The shaded region marks the excluded part of the parameter space.} 
\label{f:photonorbit}
\end{figure}

\begin{figure}[ht]
\begin{center}
\psfig{figure=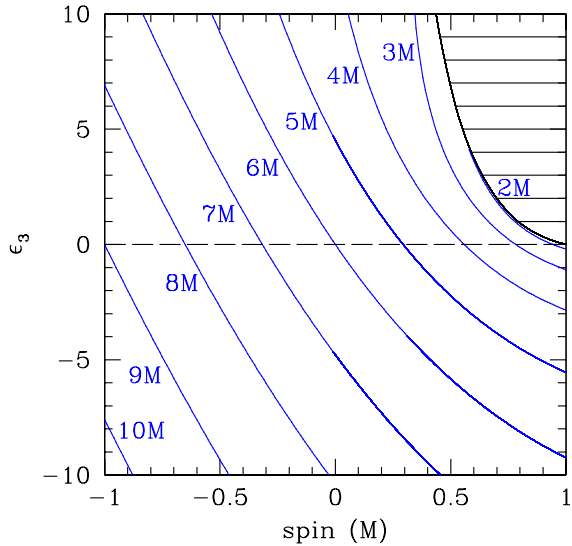,height=3.3in}
\end{center}
\caption{Contours of constant radius of the ISCO for values of the spin $-1\leq a/M \leq1$ and of the parameter $-10\leq \epsilon_3 \leq 10$. The radius of the ISCO decreases for increasing values of the spin and the parameter $\epsilon_3$. The shaded region marks the excluded part of the parameter space. The dashed line corresponds to the parameter space for a Kerr black hole.}
\label{f:iscocontours}
\end{figure}

\begin{figure}[ht]
\begin{center}
\psfig{figure=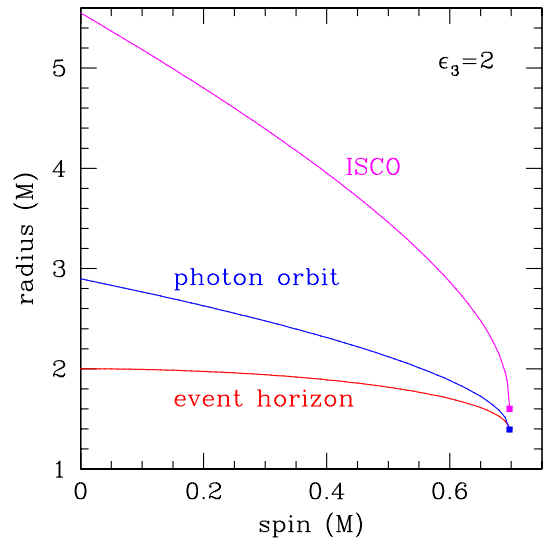,height=3.3in}
\end{center}
\caption{Equatorial radius of the event horizon, the circular photon orbit, and the ISCO as a function of the spin $a$ for a value of the parameter $\epsilon_3=2$. The event horizon and the circular photon orbit coincide at $r\approx1.39M$ at a spin of $a\approx0.697M$, while the ISCO reaches a value of $r\approx1.60M$.} 
\label{f:merging}
\end{figure}

\begin{figure}[ht]
\begin{center}
\psfig{figure=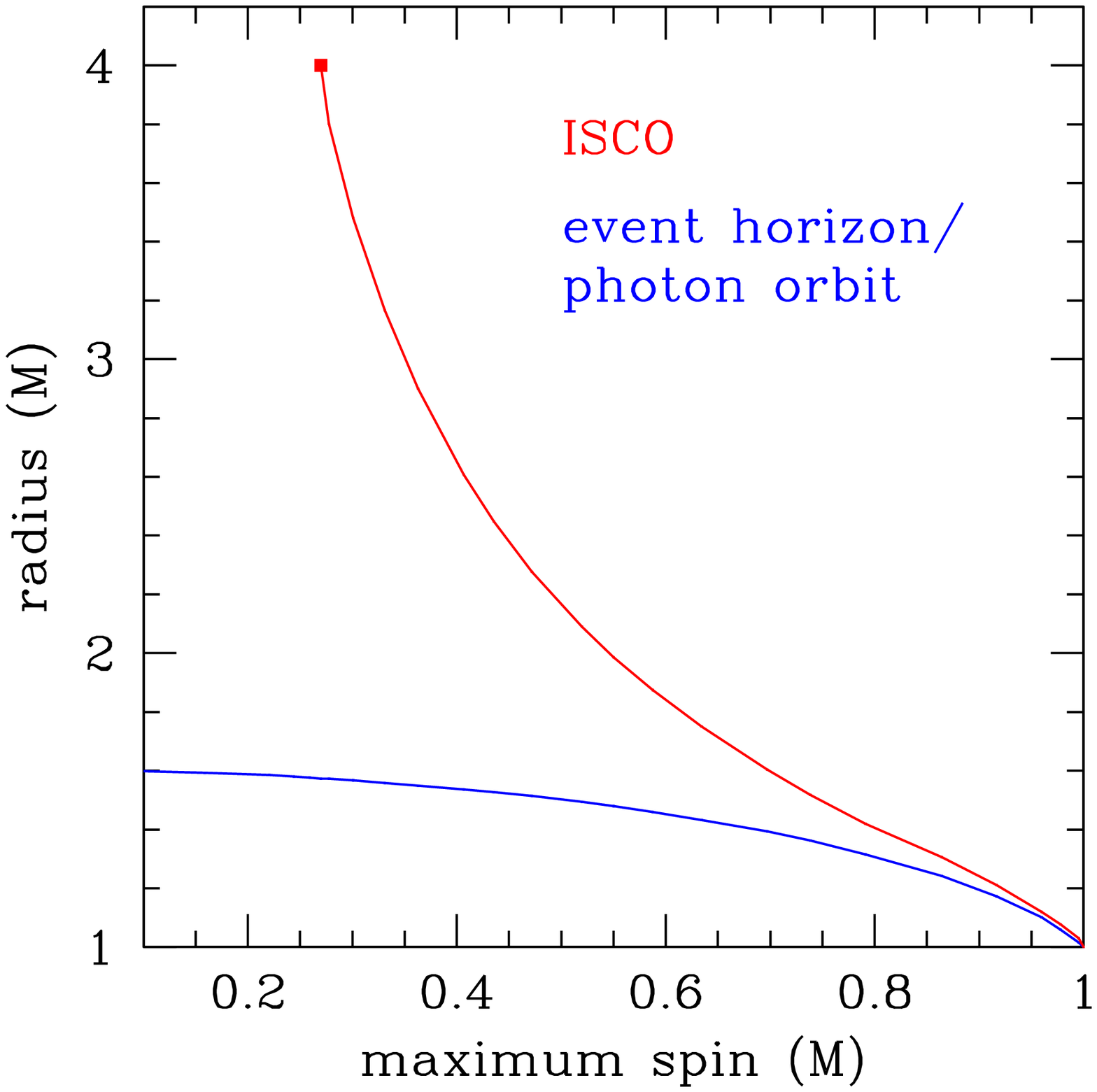,height=3.3in}
\end{center}
\caption{Radii of the (equatorial) event horizon and the prograde circular photon orbit and ISCO as a function maximum spin. The event horizon and the circular photon orbit coincide for all values of the maximum spin. The prograde ISCO at these values of the spin is located at slightly larger radii and merges with the event horizon and the circular photon orbit in the Kerr limit $a_{\rm max}=M$. For values of the maximum spin smaller than (red dot) $a_{\rm max}\approx0.270M$, multiple ISCOs occur (not shown).}\label{f:maxhorizon}
\end{figure}

\subsection{Energy and Axial Angular Momentum for a Particle on a Circular Equatorial Orbit}

Here we derive expressions for the energy $E$ and axial angular momentum $L_z$ of a particle on a circular equatorial orbit. Our derivation is similar to the ones in \cite{Bardeen73} for the Kerr metric and in \cite{PaperI} for the quasi-Kerr metric.

Since our metric is stationary and axisymmetric, there exist three integrals of the motion. For a particle with 4-momentum
\be
p^\alpha = \mu \frac{dx^\alpha}{d\tau},
\ee
these constants are its rest mass $\mu$, energy $E=-p_t$, and axial angular momentum $L_z=p_\phi$.

The Kerr metric (in Boyer-Lindquist coordinates) is of Petrov-type D, which ensures the existence of a fourth constant of the motion \cite{Carter68}. Our metric in the form given by Eq. (\ref{metric}) with the function $h(r,\theta)$ chosen according to Eq. (\ref{hchoice}) is of Petrov-type I, and the fourth constant is lost. However, thanks to the reflexion symmetry of our spacetime, equatorial trajectories are fully characterized by the rest mass, energy, and axial angular momentum alone.

We solve the equation
\be
p_\alpha p^\alpha = -\mu^2
\label{restmassconservation}
\ee
in the equatorial plane for the radial momentum and insert the constants of the motion. We obtain
\ba
&& \left(\frac{dr}{d\tau}\right)^2 \equiv R(r) \nonumber \\
&& \equiv -\frac{1}{g_{rr}} \left( g^{tt}E^2 - 2g^{t\phi}E L_z + g^{\phi\phi}L_z^2 + \mu^2 \right),
\ea
where $g_{\alpha\beta}$ is our metric given by Eqs. (\ref{metric}) and (\ref{hchoice}).

We solve the system of equations
\ba
R(r) &=& 0, \\
\frac{d}{dr}R(r) &=& 0
\ea
for the energy and axial angular momentum and find the expressions
\be
\frac{E}{\mu} = \frac{1}{r^6}\sqrt{ \frac{P_1 + P_2}{P_3} },
\label{energy}
\ee
\ba
\frac{L_z}{\mu} = && \pm \frac{1}{r^4 P_6 \sqrt{P_3}} \big[ \sqrt{M(r^3+\epsilon_3 M^3) P_5 } \nonumber \\
&& \mp 6a M (r^3 + \epsilon_3 M^3) \sqrt{P_1 + P_2} \big].
\label{angularmomentum}
\ea
In these expressions, the upper signs refer to a particle that corotates with the black hole, while the lower signs refer to a counterrotating particle. The functions $P_1$ to $P_6$ can be found in Appendix A.

In the Kerr limit, $\epsilon_3\rightarrow0$, these expressions simplify to the corresponding ones for the Kerr metric \cite{Bardeen73}
\be
\frac{E}{\mu} = \frac{ r^{3/2} - 2Mr^{1/2} \pm aM^{1/2} }{ r^{3/4}\sqrt{r^{3/2} - 3Mr^{1/2} \pm 2aM^{1/2}} }
\ee
and
\be
\frac{L_z}{\mu} = \pm \frac{ M^{1/2}(r^2 \mp 2aM^{1/2} r^{1/2} +a^2) }{ r^{3/4}\sqrt{r^{3/2} - 3Mr^{1/2} \pm 2aM^{1/2}} }.
\ee

\subsection{Innermost Stable Circular Orbit and Circular Photon Orbit}

From the expressions (\ref{energy}) and (\ref{angularmomentum}) for the energy and axial angular momentum, we derive the locations for the ISCO and the circular photon orbit. In order to obtain the ISCO, we numerically solve the equation
\be
\frac{dE}{dr} = 0.
\ee
The photon orbit occurs at the radius at which
\be
\frac{E}{\mu}\rightarrow\infty,~~~~~\frac{L_z}{\mu}\rightarrow\infty,
\ee
and the denominators in the expressions (\ref{energy}) and (\ref{angularmomentum}) vanish. Compared to the denominator of the energy $E/\mu$, the denominator of the angular momentum $L_z$ contains the additional factor $P_6$. This factor, however, has no real roots, and we can determine the radius of the photon orbit uniquely. Similar calculations have also been performed by \cite{ShibataBerti,PaperI}.

In Figures~\ref{f:isco} and \ref{f:photonorbit}, we plot, respectively, the ISCO and the circular photon orbit as a function of the spin for several values of the parameter $\epsilon_3$. The radii of the ISCO and of the circular photon orbit decrease with increasing values of the parameter $\epsilon_3$. The shaded regions mark the excluded part of the parameter space in accordance with Figure~\ref{f:maxdat}. A spacetime with an ISCO or photon orbit radius inside the shaded regions would have an open event horizon. The solid lines along the boundary of the excluded part correspond to the locations of the ISCO and the circular photon orbit, respectively, for the range of the parameter $0\leq\epsilon_3\leq10$. We do not calculate the boundary for values of the parameter $|\epsilon_3|>10$ explicitly and estimate its location in both figures by a dashed line. 

In Figure~\ref{f:iscocontours}, we plot contours of constant radius of the ISCO as a function of spin and the parameter $\epsilon_3$. The radius of the ISCO decreases for increasing values of the spin and the parameter $\epsilon_3$. The shaded region marks the excluded part of the parameter space. The dashed line corresponds to the parameter space for a Kerr black hole, which depends only on the spin.

In the Kerr metric in Boyer-Lindquist coordinates, the equatorial event horizon, the prograde circular photon orbit, and the ISCO coincide at maximum spin $a=M$ even though their proper separation is distinct. For values of the spin exceeding this upper bound, the central object is no longer a black hole. In our metric, the upper bound depends on both the spin and the parameter $\epsilon_3$ as we have already shown in Figure~\ref{f:maxdat}. Along this curve, the equatorial event horizon and the prograde circular photon orbit merge within numerical accuracy, while the prograde ISCO is located at a radius slightly outside of the event horizon.

We illustrate this behavior in Figure~\ref{f:merging}, where we plot the equatorial radius of the event horizon, the circular photon orbit, and the ISCO as a function of the spin $a$ for a value of the parameter $\epsilon_3=2$. At a spin of $a\approx0.697M$, the circular photon orbit merges with the event horizon at $r\approx1.39M$, while the ISCO reaches a value of $r\approx1.60M$. For values of the spin larger than the upper bound $a\approx0.697M$, the event horizon is no longer closed.

In Figure~\ref{f:maxhorizon}, we plot the radii of the equatorial event horizon and of the prograde circular photon orbit and ISCO as a function of the maximum spin. The event horizon and the circular photon orbit coincide for all values of the maximum spin reaching the asymptotic value $r=2.0M$ in the limit $a_{\rm max}\rightarrow0$, $\epsilon_3\rightarrow\infty$. The prograde ISCO at these values of the spin is located at slightly larger radii and merges with the event horizon and the circular photon orbit in the Kerr limit $a_{\rm max}=M$. For values of the maximum spin smaller than $a_{\rm max}\approx0.270M$ (corresponding to a value of the parameter $\epsilon_3\approx32$), the prograde ISCO is no longer unique, and a second region of stable circular orbits occurs between the origin and the ISCO separated by a gap. This region is bound by both another innermost as well as an outermost stable orbit (``OSCO''). This region lies outside of the parameter space that we consider in this paper (c.f., Figure~\ref{f:maxdat}). A similar effect in other parametric spacetimes has also been reported in Refs. \cite{PaperIII,Gair08}.

\section{Conclusions}

Thanks to the no-hair theorem, any parametric deviation from the Kerr metric in general relativity does not harbor a black hole and is often plagued with unphysical properties that have to be excluded by imposing a cutoff near but outside of the event horizon. Within general relativity, tests of the no-hair theorem that are based on observational signals originating from the vicinity of the circular photon orbit or the ISCO are, therefore, limited to the region outside of the cutoff, and, so far, it has been unknown how to include rapidly spinning black holes in such tests \cite{JP11b}.

In this paper, we constructed a Kerr-like metric of a rapidly spinning black hole, which depends on a set of free parameters in addition to the mass and spin and which reduces smoothly to the Kerr metric if all parameters vanish. We showed that this metric is stationary, axisymmetric, and asymptotically flat and argued that it describes a vacuum spacetime for a set of appropriately chosen field equations. We used the current results from Lunar Laser Ranging tests of weak-field general relativity to constrain the set of free parameters.

For the case of one additional parameter, we showed that our metric is regular and free of unphysical properties outside of the event horizon and that it can be used to describe black holes up to the maximum value of the spin $a$. For positive values of the free parameter, this upper bound is a function of the deviation and smaller than the Kerr value $a_{\rm max}=M$. Otherwise, the upper bound coincides with the Kerr limit. For values of the spin $|a|>M$ and of the parameter $\epsilon_3 < -16|a|^3 / 3\sqrt{3}$, our metric describes a superspinning black hole.

We calculated expressions for the energy and angular momentum of a particle on a circular equatorial orbit and used them to obtain the locations of the ISCO and the circular photon orbit, respectively. Both radii decrease with increasing values of the spin and the deviation parameter. At the maximum value of the spin for a given value of the deviation, the circular photon orbit merges with the event horizon within numerical accuracy as in the Kerr metric, and the ISCO is located slightly outside of the horizon.

Our metric is, thus, fully applicable in the strong-field regime arbitrarily close to the event horizon of a black hole and an ideal spacetime for astrophysical tests of the no-hair theorem that probe the immediate vicinity of black holes and that do not rely on the field equations explicitly.

We thank S. Hughes, L. Stein, S. Vigeland, and N. Yunes for many useful discussions. This work was supported by the NSF CAREER award NSF 0746549.

\appendix
\begin{widetext}
\section{Energy and Angular Momentum for a Particle on a Circular Equatorial Orbit}

In this appendix, we give explicit expressions for the functions $P_1$ to $P_6$ that occur in the expressions (\ref{energy}) and (\ref{angularmomentum}) for the energy and axial angular momentum, respectively, of a particle on a circular equatorial orbit:
\ba
P_1 = &&a^2 M r^4 \left(\epsilon_3 M^3 +r^3\right)^2 \bigg\{12\epsilon_3 a^2 M^3 \left(\epsilon_3 M^3+r^3\right)^2 \nonumber \\
&&-r^4 \left[2\epsilon_3 M^2 r^3 \left(3 r^2-8 M^2\right)+\epsilon_3^2 M^5 \left(40 M^2-48 M r+15 r^2\right)+4 r^6 (3 r-5 M)\right]\bigg\} \\
P_2 = &&2 \bigg\{2 r^4\left(r^{20} \mp M P_4\right)+M r^{12} \bigg\{ 2 r^9 \left(-12 M^2+16 M r-7r^2\right) \nonumber \\
&&+\epsilon_3 M^2 (r-2 M)^2 \left[\epsilon_3^2 M^6 (5r-12 M)-6\epsilon_3 M^3 r^3 (5 M-2 r)-3 r^6 (8 M-3r)\right]\bigg\}\bigg\} \\
P_3 = &&r^4 \left(12\epsilon_3 M^4 -5\epsilon_3 M^3 r +6 M r^3-2 r^4\right)^2-8 a^2 M \left(\epsilon_3 M^3 +r^3\right)^2 \left(5\epsilon_3 M^3 +2 r^3\right) \\
P_4 = && \sqrt{a^2 M \left(\epsilon_3 M^3 +r^3\right)^6 \left(9\epsilon_3^2 a^2 M^5+16\epsilon_3 M^3 r^4 -6\epsilon_3 M^2 r^5 +4 r^7\right) \left[a^2 \left(\epsilon_3 M^3 +r^3\right)+r^4 (r-2 M)\right]^2} \\
P_5 = &&\left(\epsilon_3 M^3+r^3\right) \bigg\{12\epsilon_3 a^6 M^3 \left(\epsilon_3 M^3 -2 r^3\right)^2 \left(\epsilon_3 M^3+r^3\right)^4+a^4 r^4 \left(\epsilon_3 M^3+r^3\right)^2 \nonumber \\
&&\left(-40\epsilon_3^4 M^{13}+40\epsilon_3^4 M^{12} r -15\epsilon_3^4 M^{11} r^2+128\epsilon_3^3 M^{10} r^3 -296\epsilon_3^3 M^9 r^4 +54\epsilon_3^3 M^8 r^5-924\epsilon_3^2 M^7 r^6 \right.\nonumber \\
&&\left.+276\epsilon_3^2 M^6 r^7 -36\epsilon_3^2 M^5 r^8 -880\epsilon_3 M^4 r^9 +304\epsilon_3 M^3 r^{10}-24\epsilon_3 M^2 r^{11}-112 M r^{12}+16 r^{13}\right) \nonumber \\
&&-2 a^2 r^8 \big[48 \epsilon_3^5 M^{17} -12\epsilon_3^5 M^{16} r -52\epsilon_3^5 M^{15} r^2+3\epsilon_3^4 M^{14} r^3 (5\epsilon_3+88)-720\epsilon_3^4 M^{13} r^4 +298\epsilon_3^4 M^{12} r^5 \nonumber \\
&& -3\epsilon_3^3 M^{11} r^6 (13 \epsilon_3+480)+516\epsilon_3^3 M^{10} r^7 +2\epsilon_3^3 M^9 r^8 -6\epsilon_3^2 M^8 r^9 (3\epsilon_3+508)+2292\epsilon_3^2 M^7 r^{10} \nonumber \\
&& -628 \epsilon_3^2 M^6 r^{11}+12\epsilon_3 M^5 r^{12} (5\epsilon_3-134)+1188\epsilon_3 M^4 r^{13} -296\epsilon_3 M^3 r^{14} +24 M^2 r^{15} (\epsilon_3-9)+120 M r^{16} \nonumber \\
&&-16r^{17}\big]-r^{14} \left(\epsilon_3 M^3 +6 M r^2-2 r^3\right)^2 \nonumber \\
&&\left(96\epsilon_3^2 M^7 -76\epsilon_3^2 M^6 r +15\epsilon_3^2 M^5 r^2+72\epsilon_3 M^4 r^3 -44\epsilon_3 M^3 r^4 +6\epsilon_3 M^2 r^5 +12 M r^6-4 r^7\right)\bigg\} \nonumber \\
&&\mp 4 P_4 \left[ a^2 \left(\epsilon_3 M^3 -2 r^3\right)^2 \left(\epsilon_3 M^3+r^3\right)+6\epsilon_3 M^3 r^5 \left(\epsilon_3 M^3 +6 M r^2-2 r^3\right)\right] \\
P_6 = && -\epsilon_3^2 M^6 -6\epsilon_3 M^4 r^2 +\epsilon_3 M^3 r^3 -6 M r^5+2 r^6
\ea

\end{widetext}

\end{document}